\documentclass{emulateapj}
\usepackage{amsmath}
\usepackage{color}

\newcommand{\Msun}{M_\odot}
\newcommand{\sci}[1]{\times10^{#1}}
\newcommand{\fig}[1]{Figure \ref{fig:#1}}
\newcommand{\figs}[2]{Figures \ref{fig:#1} and \ref{fig:#2}}

\newcommand{\eqn}[1]{Equation (\ref{eqn:#1})}

\newcommand{\lcdm}{$\Lambda$CDM}
\newcommand{\kms}{km\,s$^{-1}$}

\newcommand{\degrees}{^\circ}

\newcommand{\waynerev}{}

\newcommand{\waynerevv}{}

\newcommand{\waynerevvv}{}

\newcommand{\referee}{}

\defcitealias{hex}{L11}
\defcitealias{ngan14}{NC14}

\shorttitle{Tidal streams in high-resolution halo}
\shortauthors{Ngan, Bozek, Carlberg, et al}

\begin{document}


\title{Simulating tidal streams in a high-resolution dark matter halo}


\author{Wayne Ngan$^1$, Brandon Bozek$^{2,3}$,
Raymond G. Carlberg$^1$, Rosemary F. G. Wyse$^3$, Alexander S. Szalay$^3$, Piero Madau$^4$}

\affil{$^1$ Department of Astronomy and Astrophysics, University of Toronto, Toronto, ON, M5S 3H4, Canada \\
$^2$ Department of Astronomy and Joint Space-Science Institute, University of Maryland, College Park, MD, 20742, USA\\
$^3$ Department of Physics and Astronomy, The Johns Hopkins University, Baltimore, MD, 21218, USA \\
$^4$ Department of Astronomy and Astrophysics, University of California, Santa Cruz, CA, 95064, USA}
\email{ngan@astro.utoronto.ca}

\begin{abstract}

We simulate tidal streams in the presence \waynerev{and} absence of substructures inside
the \waynerev{zero-redshift snapshot of} the Via Lactea II (VL-2)
simulation. A halo finder is used to remove and isolate the subhalos found inside the high-resolution
dark matter halo of VL-2, and the potentials for both the main halo and all the subhalos are constructed
individually using the self-consistent field (SCF) method. This allows us to make direct comparison
of tidal streams between a smooth halo and a lumpy halo without assuming idealized profiles or triaxial
fits. We simulate \waynerevv{the kinematics of a} star cluster
starting with the \waynerevv{same} orbital position but two different velocities. Although these two orbits
are only moderately eccentric and have similar apo- and pericentric distances, we find that \waynerevv{the} two
streams have very different morphologies. We conclude that our \waynerev{model of the potential} of VL-2
can provide insights about tidal streams that have not been explored by previous studies using idealized or
axisymmetric models.

\end{abstract}





\section{Introduction}

In the \lcdm\ hierarchical structure formation model, smaller structures form first and \waynerev{merge}
to form larger structures \waynerev{\citep{blumenthal84, davis85, bardeen86}}.
This process is not complete; many of the smaller structures \waynerev{survive until
today to become substructures within larger structures like the Milky Way}. This has been \waynerev{demonstrated by}
high-resolution simulations
of Milky-Way-sized dark matter halos, which are populated by numerous subhalos
\waynerevv{\citep{diemand07,diemand08,vl2,aquarius,ghalo,zemp_review,gao11}}.
On the other hand, the observed abundances of satellite
galaxies around the Milky Way or M31 \waynerev{are} much lower than the abundances of
subhalos predicted by simulations. \waynerev{This disagreement between observation and theory
is commonly known as the missing satellite problem \citep{klypin99,moore99,strigari07}.}



Tidal streams, or remnants of stellar structures as they are tidally disrupted by a massive host, 
\waynerev{are} powerful probes for the potential in the the Milky Way
\waynerev{\citep[e.g.,][]{johnston98, helmi03, law05}}. If subhalos exist, they should interact
with the stream dynamically and leave detectable signatures along the stream. \citet{ibata02} first 
simulated a globular cluster in idealized galaxy potentials \waynerev{both with and without subhalos}
and found that the resulting streams had
\waynerev{distinguishable kinematics when subhalos were present}. 
\referee{More recently, both idealized simulations \citep{yoon11,carlberg12} and self-gravitating simulations
\citep{erkal14,ngan14}} found that a close encounter between a subhalo and \referee{a} stream
can produce a ``gap'' which can be observed as an underdensity of stars along the stream.
\waynerev{In particular, \citet{pal5gaps}, \citet{carlberg12}, and \citet{gd1gaps} analytically derived the gap formation rate
as an observable quantity, so the number of gaps can be a powerful probe for the abundance
of subhalos.}

Detailed observations of densities along \waynerevv{stellar} streams show that streams have many longitudinal underdensities
\citep{m31gaps, pal5gaps, gd1gaps}, but some studies have \waynerevv{found} that intrinsic mechanisms such as epicyclic
overdensities \citep[EOs;][]{kupper08, kupper10, kupper12} and Jeans instabilities \citep{cq11} can also 
produce gap-like features which can be confused with gaps \waynerevv{induced} by subhalo perturbations. \citet[hereafter NC14]{ngan14}
\waynerev{\referee{simulated streams} in identical orbits with and without subhalos}
and concluded that EO gaps can be distinguished from subhalo gaps \waynerev{by the locations and the size distribution of the gaps.}

All of the above simulations share the same limitation --- idealized dark matter halo profiles.
\citet{nfw, navarro04, navarro10} \waynerev{showed} that dark matter halos can be described by universal profiles
such as Navarro-Frenk-White and Einasto profiles, and the former is used by some simulations mentioned above
because of its simple form. However, idealized profiles are spherically averaged best-fit results, which
do not capture the shape of the halo. \waynerev{Both N-body simulations \citep{js02,zemp09} and the results inferred from
observations of the Milky Way \citep{lmj}}
show that halos are not spherical. \citet{sv08} simulated the disruption of a satellite galaxy and its resulting stream
in a \waynerev{flattened} potential \waynerev{both} with and without substructures,
\waynerev{and they found that even though substructures make
the stream more clumpy, the shape of the halo has a larger effect on the satellite's disruption than 
substructures have.} Nevertheless, their underlying halo potential still
follows an idealized profile, and their stream progenitor is much more massive than the streams
we consider in this study. \waynerev{In this study, we focus on the disruption of a globular cluster
in a realistic potential of a Milky-Way-sized
dark matter halo in both the presence and absence of substructures, without using idealized profiles in either case.
The realistic potential is constructed from the zero-redshift snapshot of the Via Lactea II (VL-2) simulation, which 
simulated the formation of a Milky-Way-sized dark matter halo using more than $10^9$ particles (or about $4.1\sci{3}\Msun$
per particle) in the \lcdm\ cosmological context starting from \waynerevv{redshift} $z=104$ \citep{diemand08}.}

The goal of this study is to demonstrate the difference in the appearance of tidal streams in two cases:
(1) a ``smooth'' dark matter halo without subhalos, \waynerevv{and} (2) a ``lumpy'' dark matter halo
\waynerevv{with the amount of substructures expected in a \lcdm\ cosmology.} Our simulation
parameters are chosen so that the resulting streams would be comparable to GD-1 \citep{gd1}, a dynamically
cold and narrow stream observed in the Milky Way. Our simulations are not meant to be physical models
of GD-1, though, because VL-2 is not a physical model of the the Milky Way, and we are only constructing
a time-independent potential using \waynerev{one} snapshot at \waynerev{redshift zero} of VL-2. \waynerev{The details of} orbital or stream
dynamics in the VL-2 potential \waynerev{are beyond the scope of this study}, as they are complicated topics that warrant
much more focused studies \waynerev{in the future}.
This study is meant to present the method to construct realistic dark matter halo potentials 
\waynerevvv{in the presence and absence of substructures requiring neither idealized profiles nor triaxial fits.
This allows us to isolate and investigate the effects that substructures have on a GD-1-like stream in a more realistic setting
than in \citetalias{ngan14}.}

\waynerev{Since the tidal disruption
of a star cluster is an ongoing process that lasts as long as the formation of the dark matter halo itself,
a redshift-dependent potential using all available snapshots of VL-2 is our \waynerev{goal in the future}. 
\waynerevvv{\citet{bonaca2014} approached this by using a ``live'' halo, which essentially resimulated
VL-2 but approximated the formation of streams. Another advantage of a live halo is that it can respond to
an external system in order to account for effects such as dynamical friction.
On the other hand, a live halo may be computationally expensive depending on its resolution.}
\waynerevv{We did not use a live halo because our goal is not to rerun VL-2, but to
construct a realistic model of the existing VL-2 halo at arbitrarily high accuracy and use this model as a background
for our own \waynerevvv{N-body} simulations.}
Moreover, our streams' masses are low enough that they will not affect
the evolution of the halo, and dynamical friction is negligible \citepalias{ngan14}.

This paper is organized as follows. Section \ref{sec:method} describes the method of using the
self-consistent field (SCF) method to construct
the potential of VL-2, which is done after identifying the subhalo particles using a halo finder. Section 
\ref{sec:stream_simulations} summarizes the parameters for our simulations such as our choice of orbits.
Section \ref{sec:results_and_discussion} is the key part of this study, which presents the goodness of the SCF,
as well as the details of the simulated streams. Section \ref{sec:summary_and_conclusion}
\waynerevv{summarizes} our results.



\ifdefined \FORMATBYHAND
	\vspace{0.5in}
\fi

\section{Method}
\label{sec:method}

\subsection{SCF Method}

Originally developed to compute collisionless N-body dynamics for galaxies \citep{scf}, the SCF method 
solves the gravitational Poisson equation by basis decomposition. This method is \waynerev{optimized for dark matter
halos} since its lowest-order basis function is already an idealized profile, and the higher-order
basis functions can be used to describe the radial and angular deviations from this idealized profile. The 
basis functions are bi-orthonormal and complete, so they can model any matter distribution \waynerev{as the decomposition order
increases.}

The approach to model the gravitational potentials in the snapshots of existing simulations of dark matter halos
has been studied extensively by \citet[hereafter L11]{hex}. Even with moderate decomposition orders, \citetalias{hex} was
able to recover much of the detailed dynamics inside the halos of the Aquarius simulations \citep{aquarius}.
We follow their approach to compute the forces in the halo of the VL-2 simulation
at redshift zero. In this section we briefly summarize the method.

The density $\rho(\mathbf{r})$ and gravitational potential $\Phi(\mathbf{r})$ are related by the Poisson equation
$\nabla^2 \Phi=4\pi G \rho$. Given a simulation snapshot that contains $\rho$, SCF solves the Poisson equation by decomposing
$\Phi$ and $\rho$ such that
\begin{eqnarray}
	\Phi(\mathbf{r}) = \sum_{nlm} A_{nlm} \Phi_{nlm}(\mathbf{r})
	\label{eqn:phi} \\
	\rho(\mathbf{r}) = \sum_{nlm} A_{nlm} \rho_{nlm}(\mathbf{r})
\end{eqnarray}
where $\mathbf{r}\equiv(r,\phi,\theta)$ is expressed in spherical coordinates for convenience.
We follow the derivation of \citet{scf} which used the Hernquist profile \citep{hernquist} where
$\Phi_{000}\equiv-1/(1+r)$ and $\rho_{000}\equiv 1/[2\pi r (1+r)^3]$ are the zeroth-order basis functions.
The general basis functions can be written as
\begin{eqnarray}
	\Phi_{nlm} \propto - \frac{r^l}{(1+r)^{2l+1}} C_n^{(2l+3/2)}(\xi) Y_{lm}(\theta,\phi)
	\label{eqn:phi_nlm} \\
	\rho_{nlm} \propto \frac{r^l}{r(1+r)^{2l+3}} C_n^{(2l+3/2)}(\xi) Y_{lm}(\theta,\phi)
\end{eqnarray}
where $Y_{lm}$ are the spherical harmonics, $C_n^{(\alpha)}(\xi)$ are the Gegenbauer polynomials, and
$\xi\equiv(r-1)/(r+1)$. Taking advantage of the bi-orthonormality of the basis functions, each $A_{nlm}$
can be computed based on a given $\rho(\mathbf{r})$ by
\begin{equation}
	A_{nlm}\propto\int \rho(\mathbf{r})\Phi_{nlm}^*(\mathbf{r}) d\mathbf{r}.
\end{equation}
The accelerations $-\nabla\Phi$ can be obtained by differentiating \eqn{phi} analytically.

In practice, a snapshot from an N-body simulation contains $\rho(\mathbf{r})$ which is represented by discrete particles,
and the computations are much more easily done in real space. We refer the
reader to \citet{scf} for a straightforward recipe to compute accelerations given a list of particle
positions and masses. \waynerev{A} noteworthy feature of the SCF method is that all accelerations are
proportional to the terms
\begin{equation}
	\sum_{k=1}^N m_k \tilde\Phi_{nl}(r_k) P_{lm}(\cos\theta_k) \left[ \begin{array}{c} \cos(m\phi_k) \\\sin(m\phi_k) \end{array} \right]
	\label{eqn:parallel_sum}
\end{equation}
where $m_k$ and $(r_k, \theta_k, \phi_k)$ are the mass and position of the $k$th particle, $\tilde\Phi_{nl}$
is the radial part of \eqn{phi_nlm}, and $P_{lm}$ are the Legendre polynomials.
These linear summations over all $N$ particles can be easily parallelized and precomputed given $(n,l,m)$.
The decomposition is truncated at orders $n_{max}$ and $l_{max}$ (and $m$ goes from 0 to $l$ in real space), 
so both terms in \eqn{parallel_sum} are evaluated
\begin{equation}
    \frac{n_{max}(l_{max}+1)(l_{max}+2)}{2}
    \label{eqn:number_of_terms}
\end{equation}    
times. After \eqn{parallel_sum} is precomputed, the accelerations can be obtained analytically.

A detailed strategy to parallelize the SCF method has been proposed by \citet{scf_parallel} to compute
N-body dynamics using the SCF method itself. Note that our problem is even simpler because we are only
extracting the accelerations from an existing N-body snapshot, and the accelerations do not need to be
propagated back to the bodies. The extracted accelerations serve as \waynerev{``external forces''} for each particle
of our stream simulations.


\waynerev{SCF offers a convenient way to account for redshift dependence.
Since the model of one static potential is entirely represented by
the set of coefficients $A_{nlm}$ in \eqn{phi}, these coefficients can simply be interpolated in redshift.
\citetalias{hex} showed that the time variation of the first few coefficients at the lowest orders are relatively small,
but higher-order variations may be more difficult to capture. However, this is unlikely to be a problem for us because
we account for subhalos separately from the smooth halo. We defer the construction of a redshift-dependent
potential to future studies.}


\subsection{Subhalo Finding}

We use the {\sc Amiga Halo Finder}\footnote{\url{http://popia.ft.uam.es/AHF/Download.html}} (AHF) to identify bound
structures in the high-resolution region of VL-2. The most massive halo, hereafter the
``main halo,'' is the halo that hosts the subhalos and the tidal streams of interest in this study.
Many particles in the main halo are also bound into additional levels of substructures. In the first level
there are 11,523 ``subhalos'' that consist of at least 150 particles.
\waynerev{The minimum halo mass threshold is chosen such that the average peak circular velocity for those
halos ($V_{max}\sim2.5$ \kms) approximately corresponds to the peak circular velocity below which the subhalo
abundance is suppressed by numerical limitations \waynerevv{\citep{diemand08, diemand11}}.
This threshold choice allows for a robust sampling of the
subhalo mass function.}

In addition to each subhalo's particle membership, AHF also automatically computes their intrinsic properties such as
positions at the minimum potential,
bulk velocities, scale radii, virial radii, etc. Table \ref{table:halo_masses} summarizes the facts about our main halo,
as well as the most and least massive subhalos.
Note that the total number of particles and total mass of the main halo include all the particles of the subhalos,
as returned by AHF. As we discuss in the following section, we remove the subhalo particles from
the main halo, so the mass of the main halo in our simulations is slightly \waynerevv{lower} than the value stated in Table
\ref{table:halo_masses}. 

%
%

\begin{deluxetable}{llllll}
    	\tablecaption{Main Halo and Subhalos in the $z=0$ Snapshot of VL-2.}
    \tablehead{
	Halo		&	$N$	&	$M$			&	$r_s$	&	$r_{vir}$\\
	\		&	\	&	($\Msun$) 	&	(kpc) 	&	(kpc)}
	\startdata
	{Main halo}               &   $4.46\sci{8}$   &    $1.89\sci{12}$   &   60.4    &    400 \\
	{Most massive subhalo}    &   $1.15\sci{6}$   &    $4.69\sci{9}$    &   6.43    &    54.2 \\
	{Least massive subhalo}   &    150             &    $6.15\sci{5}$    &   0.411   &    2.75
	\enddata
	\tablenotetext{}{
	The total number of particles ($N$), total masses ($M$), scale radii ($r_s$), and
	virial radii ($r_{vir}$) of three \waynerevv{nominal} halos extracted by the halo finder, which found
	a total of 11,523 subhalos inside the virial radius of the main halo. 
	\label{table:halo_masses}}
\end{deluxetable}

\subsection{Modeling the smooth halo}

In order to obtain a smooth halo from VL-2 that is a \lcdm\ simulation, we remove
all the subhalo particles from the main halo, leaving ``voids'' that were previously occupied by subhalos. 
When we construct the smooth halo's potential, we use a low-order decomposition that only captures the overall
shape of the main halo, but not the lumpiness due to the removed subhalos.

For convenience, for the smooth halo we only consider the cases where $n_{max}=l_{max}$,
which we will refer to as the decomposition ``order'' collectively. 
\figs{force_errors_10}{force_errors_20} show the accuracies of constructing the force field of the smooth
halo using orders 10 and 20 by comparing against the force field that is obtained by directly summing up the force contributions from all
the particles in the smooth halo. The force fields using both SCF and direct summation are computed in
94,747 randomly distributed positions inside a spherical shell of $r=[15,40]$ kpc in thickness where our stream will be orbiting.

There are 660 and 4620 terms in the decompositions using orders 10 and 20, respectively (\eqn{number_of_terms}).
\fig{force_error_distributions} shows that the \waynerev{improvements in accuracy} by increasing the order from 10 to 20 are marginal.
For the purpose of this study, it is not necessary to model VL-2 with the highest possible accuracy. Our goal is to
study the effects that a dark matter halo has on tidal streams in general, but not to model any particular objects that
have been observed.

\waynerev{One alternative way to construct a smooth halo is to perform a decomposition with similar orders on all the
main halo particles without removing the subhalo particles. We ran a similar analysis to \figs{force_errors_10}{force_errors_20}
using all particles, and we found that compared to having removed the subhalo particles, using all particles gave similar
error levels except at positions near the subhalos, where the errors would typically be a few percent higher. This was 
expected because decompositions at orders 10 or 20 cannot model the forces inside the subhalos that
were contributing to the directly summed forces. Removing the subhalo particles allowed us to confirm that
our force errors were within 1\% inside the smooth halo alone.}

\begin{figure}
    \includegraphics[width=3.5in]{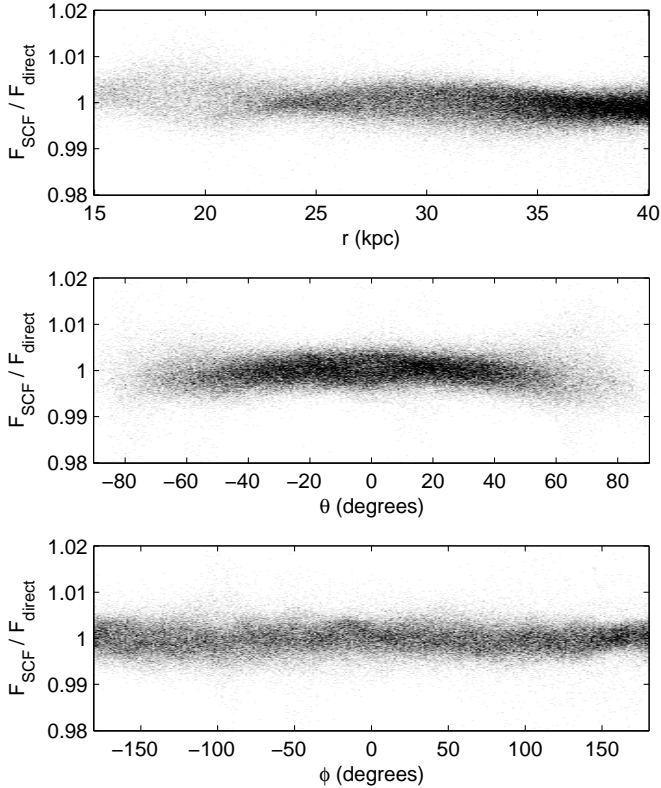}
    \caption{Comparing the magnitude of accelerations obtained from SCF ($n_{max}=l_{max}=10$) against directly summing up
    the accelerations from all particles in the smooth halo. The accelerations are measured at 94,797 randomly distributed positions inside
    a spherical shell of $r=[15,40]$ kpc. Top, middle, and bottom panels are the errors plotted against the
    $(r,\theta,\phi)$ in spherical coordinates of the test positions. The accelerations are accurate to $\sim 1 \%$ level
    \waynerev{everywhere}.}
    \label{fig:force_errors_10}
\end{figure}
\begin{figure}
    \includegraphics[width=3.5in]{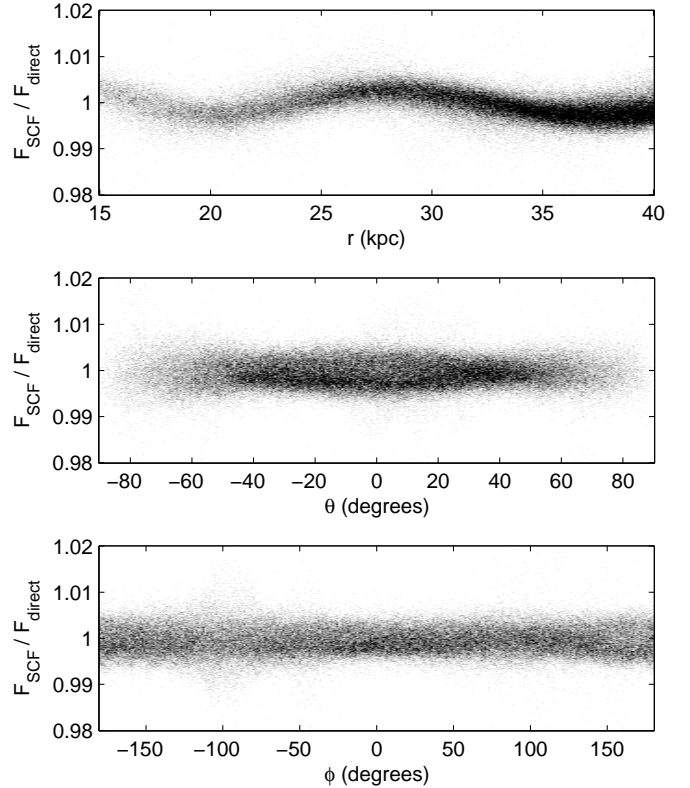}
    \caption{Same as \fig{force_errors_10}, but with $n_{max}=l_{max}=20$. \waynerev{The top panel shows some 
    correlation between the force error and radial position. This correlation is expected from basis functions whose
    radial components are polynomials. Nevertheless, the errors remain at the 1\% level, so the correlation is not a 
    concern.}}
    \label{fig:force_errors_20}
\end{figure}
\begin{figure}
    \includegraphics[width=3.5in]{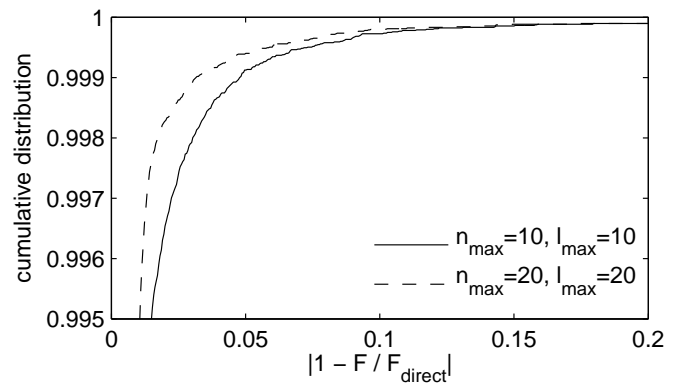}
    \caption{The cumulative distribution of the error at all the points in \figs{force_errors_10}{force_errors_20}.
    The $\sim 1\%$ level error is achieved \waynerev{more than 99\% of the time in both orders}.}
    \label{fig:force_error_distributions}
\end{figure}

\subsection{Adding Subhalos}
\label{sec:adding_subhalos}

For the \waynerev{lumpy} halo, we now add the forces by the subhalos back into the smooth halo.
We decompose each subhalo using $n_{max}=4,l_{max}=0$.
In addition to being simpler to compute, we prefer not to decompose subhalos at higher orders because some low-mass subhalos
can have only a few hundred particles, so a low $n_{max}$ and spherically symmetric decomposition can avoid unphysical
clumpiness in each subhalo. Also,
subhalos' effects on tidal streams are expected to be of short duration, so it is not necessary to model individual subhalos
in detail.


\waynerevv{Adding subhalos separately to the smooth halo provides two advantages. First of all,}
we can control the mass range of subhalos that
are present in the galaxy. This can be used to test theories such as warm dark matter that suppress the formation
of low-mass subhalos. Also, each individual subhalo can orbit freely around the smooth halo, so their 
encounters with tidal streams can be modeled realistically.

Adding subhalos separately introduces extra mass to the smooth halo when comparing
\waynerev{lumpy} and smooth halos. Our full range of subhalos have masses $1.067\sci{11}\Msun$ in total, which is 
about 6\% of the mass of the smooth halo. In our simulations, \waynerev{we only use a subset of subhalos that
(a) have pericentric distances of less than 40 kpc and (b) are less massive than $10^8\Msun$. The reason for
requirement (a) is that subhalos that do not approach the stream will have minimal effects on our streams with apocentric
distances at 30 kpc in their orbits. The reason for requirement (b) is that encounters with massive subhalos
do not leave interesting signatures. As shown in \citet{yoon11}, massive subhalos are rarer than low-mass subhalos, so 
massive subhalos mostly influence a stream by distant encounters that do not cause gaps in the stream. On the other hand,
when a massive subhalo makes a close encounter with a stream, the effect is likely catastrophic to the stream as
opposed to leaving small gaps that have been observed \citep{pal5gaps,gd1gaps}. A separate study for the effects of massive
subhalos on streams in a realistic potential is currently under way. For this study, after imposing requirements (a) and (b),}
the \waynerev{lumpy} halo has 3808 subhalos in
our simulations, which have \waynerev{a total mass of} $1.607\sci{10}\Msun$, or about $1\%$ of the mass of the 
smooth halo. Therefore, the streams simulated with and without subhalos will \waynerev{only} have slightly
different orbits, \waynerev{and the difference can be safely ignored for our comparisons.}



\section{Stream simulations}
\label{sec:stream_simulations}

\subsection{Progenitor and Orbit}

The streams' progenitor is a star cluster \waynerev{following a King profile} with core size $0.01$ kpc and 
$w=4.91$, \waynerev{where $w$ can be thought of as the ratio between the depth of the potential and central
velocity dispersion of the cluster}. Its total mass is $4.29\sci{4}\Msun$, \waynerev{central velocity dispersion is $2.4$ \kms},
and tidal radius is $0.103$ kpc \waynerev{(identical to the one used in \citetalias{ngan14})}. We use $N=1,000,000$ particles
for the the main results of this study. \waynerevv{Furthermore}, as argued in \citetalias{ngan14}, the number of particles in the stream
may \waynerev{be higher than} the number of stars in the observed streams.
\waynerev{Whether the effects we present in the following sections
can be found in existing data requires more careful investigation, which is beyond the scope of this study.}

We simulate the star cluster with two orbits shown in \fig{orbits} -- both start at $(x,y,z)=(30,0,0)$ kpc, but one orbit with 
$(v_x,v_y,v_z)=(0,120,0)$ \kms\ \referee{(``Orbit 1'')}, and the other with $(0,41.0,113)$ \kms\ \referee{(``Orbit 2'')}, which
is the former with an initial velocity inclined at $70\degrees$ from the $xy$-plane. The VL-2 main halo has not been aligned
with any axis, so the directions of the initial velocities are arbitrarily chosen to
\waynerev{explore the main halo as much as possible. Although neither orbit is confined in any
orbital planes, both are tube orbits with apo- and pericentric distances at roughly $\sim30$ kpc and $\sim17$ kpc,
which are chosen to be similar to the inferred values of the GD-1 stream \citep{gd1,willett09}}.

\begin{figure}
    \includegraphics[width=3.5in]{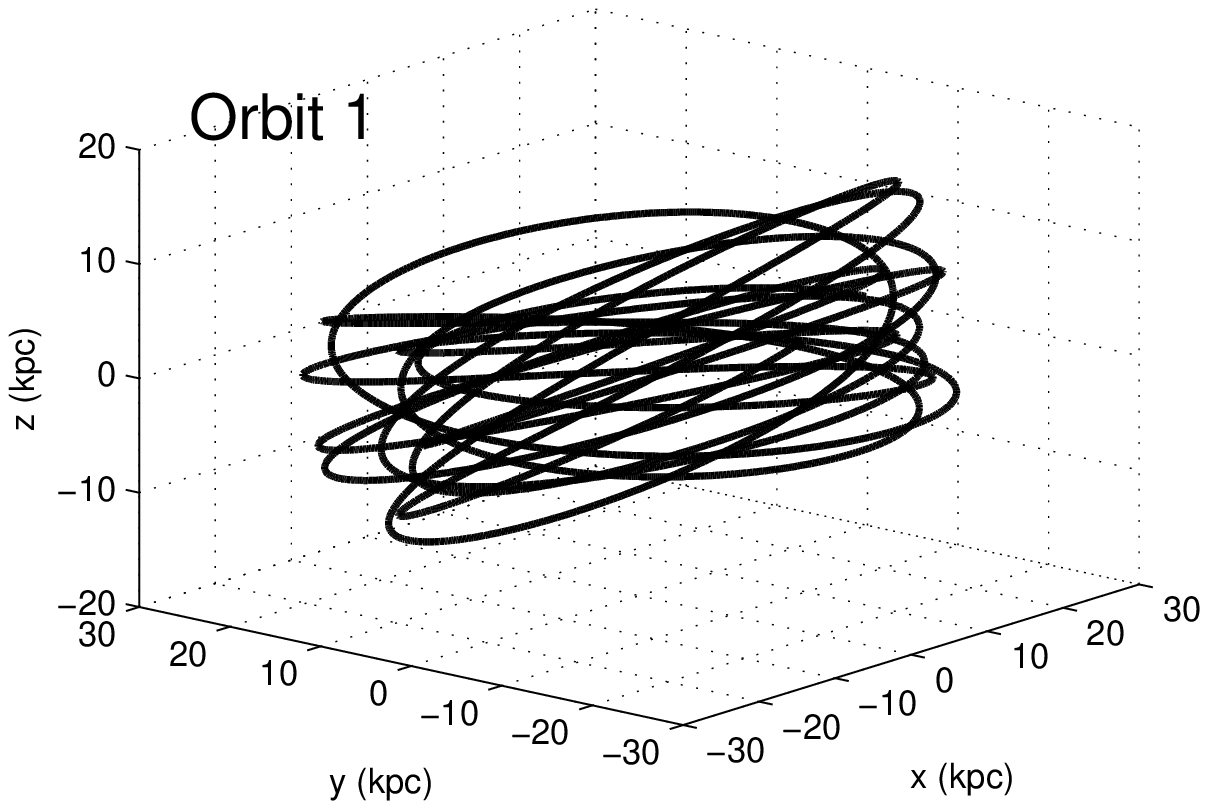} \\
    \includegraphics[width=3.5in]{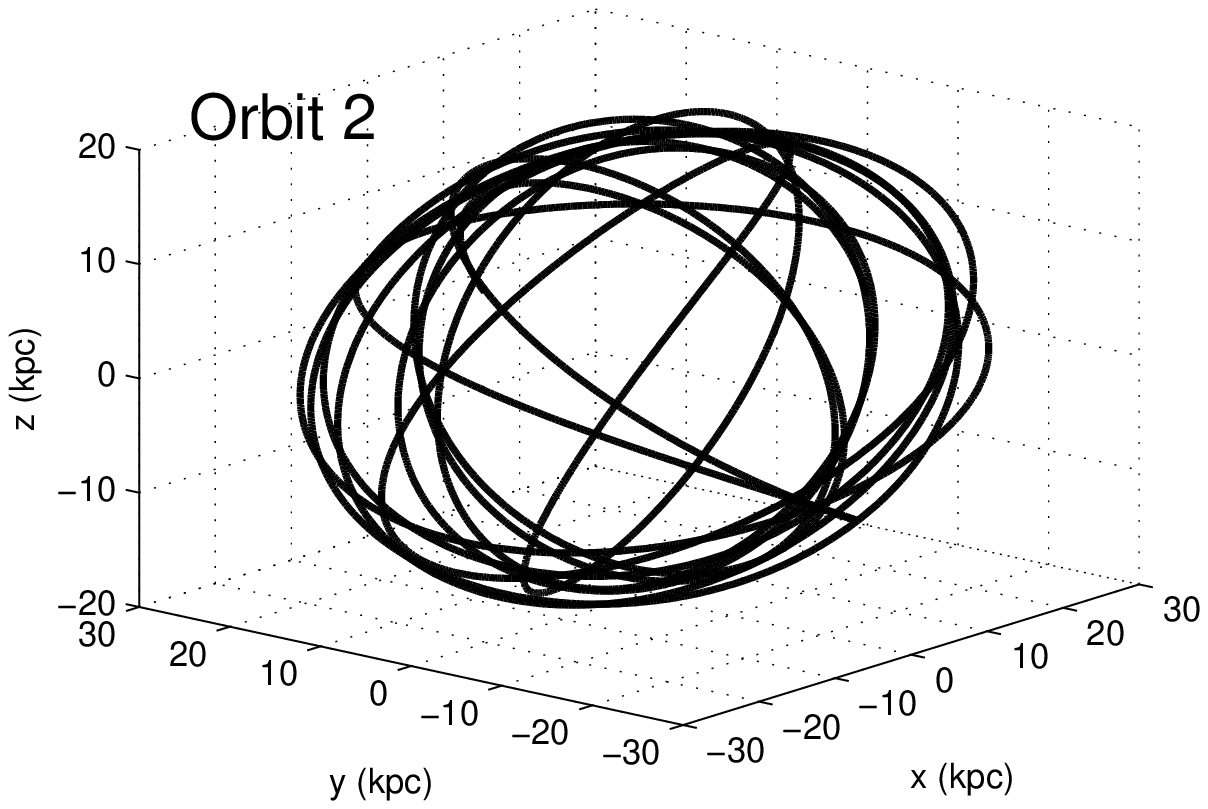}
    \caption{Two stream orbits in an SCF potential \waynerev{of the smooth halo}
    using order 10 with initial positions at $(x,y,z)=(30,0,0)$ kpc, but with
    $(v_x, v_y, v_z) = (0, 120, 0)$ \kms\ (\referee{``Orbit 1''}; top panel) and $(0,41.0,113)$ \kms\ (\referee{``Orbit 2''}; bottom panel).
    Both are tube orbits with apo- and pericenters that roughly correspond to the \waynerev{inferred}
    values of GD-1.}
    \label{fig:orbits}
\end{figure}

\subsection{Simulation Details}

The SCF decomposition is performed using the procedure exactly described in \citet{scf}. Since we consider
decomposition orders that are much lower than the number of particles in VL-2, it is not necessary to 
apply a smoothening kernel to the VL-2 particles when computing the decomposition coefficients because a
low-order decomposition, by construction, cannot capture the granularity of the VL-2 particles.

Our N-body simulations of streams are computed using {\sc Gadget-2} \citep{gadget}, which is available to
the public\footnote{\url{http://www.mpa-garching.mpg.de/gadget/}}. We modify {\sc Gadget-2} so it 
uses the precomputed SCF decomposition coefficients to construct the external accelerations, which are added
to the particles in our stream simulations after the stream particles' N-body forces have been computed.
We impose a maximum time step of 1 Myr and softening of 5 pc for the particles so that they are essentially
collisionless.

\section{Results and Discussion}
\label{sec:results_and_discussion}

\subsection{Decomposition Order}

\subsubsection{Orbital Convergence}

The basis functions in the SCF method are complete, which means that the higher the decomposition order, the more
accurate the constructed model will be. It is crucial to 
understand the decomposition order necessary for our purposes. Neither VL-2 nor our stream simulations are
meant to be physical models of any observations, so it is not necessary for us to model the VL-2 main halo to the
highest possible accuracy with a high decomposition order. In fact, since we have removed subhalos from the main halo,
modeling these ``voids'' in the smooth halo may result in spurious forces. On the other hand, if the 
decomposition order is too low, then it compromises the value in using VL-2, which features shapes and profiles
missing in idealized profiles. We require a decomposition order that is high enough to capture the shape and
profile of the VL-2 smooth halo, but not high enough to capture the granularity of its substructures
(the contribution by substructures will be added in after the smooth halo has been modeled).

The radial \waynerev{coordinates} of the two stream orbits as functions of time for various decomposition orders are shown
in \fig{orbits_radial}. Both orbits have converged to within a few percent everywhere \waynerev{for 10 Gyr}. This
is expected from \figs{force_errors_10}{force_errors_20}, which show that the forces almost everywhere are accurate to
$\sim 1\%$ even at only order 10. Therefore, a decomposition order 10 is sufficient for our purposes.
Using order 10 has the advantage that subhalo ``voids'' will not be captured. There are more than 10,000 subhalos
found within the virial radius of the main halo, and the granularities of these subhalos cannot be modeled by
polynomials of only the 10th degree in either the radial or angular components of SCF basis functions.

\citetalias{hex} showed that decompositions lower than order 10 give significant
errors in the region \waynerev{less than a few kiloparsecs away from the halo's center}, so \citetalias{hex} adapted order 20 
\waynerev{to suppress those errors. That region is not relevant for our streams because they orbit at galactocentric
distances well beyond a few kiloparsecs.}
Nevertheless, our convergence results agree with \citetalias{hex}, which showed that the acceleration errors compared
to direct summation are at $\sim 1\%$ beyond 17 kpc, and that increasing from order 9 to 19
provides minimal gain at those regions.

\begin{figure}
    \includegraphics[width=3.5in]{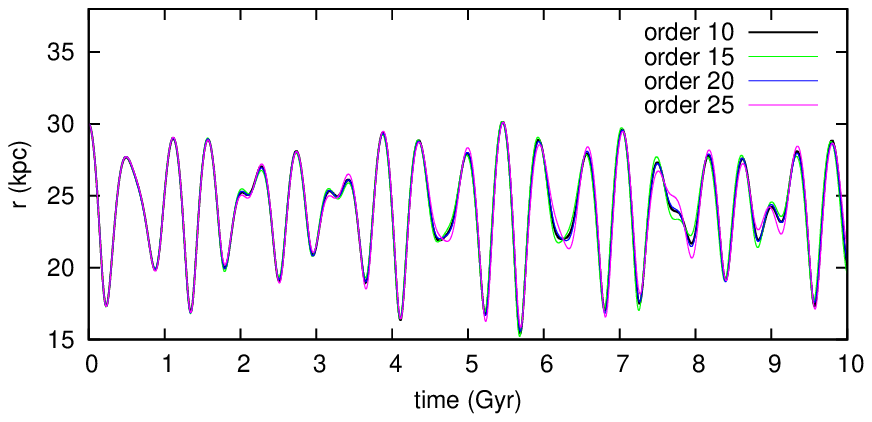} \\
    \includegraphics[width=3.5in]{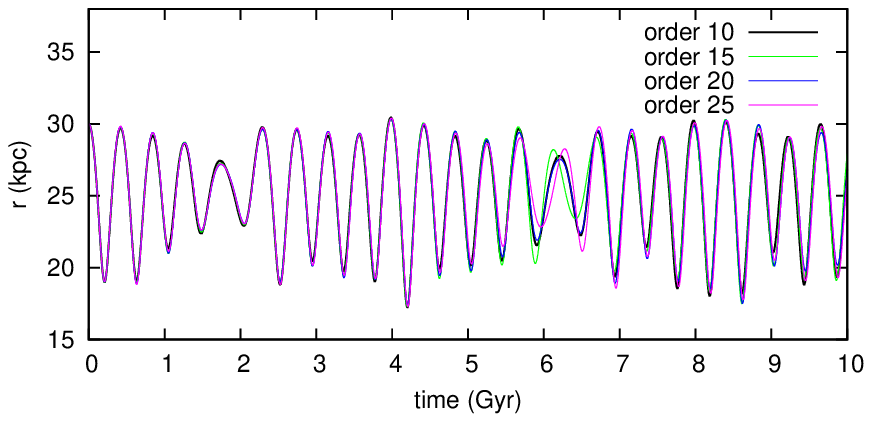} \\
    \caption{Radial \waynerev{coordinates} of two stream orbits shown in \fig{orbits}. \referee{Top and bottom 
    panels show Orbits 1 and 2, respectively.} These two plots show that their orbits
    have converged at order 10.}
    \label{fig:orbits_radial}
\end{figure}

\subsubsection{Stream Density Convergence}

As shown in \citetalias{ngan14}, signatures of subhalos can be detected in the linear density, \waynerev{or number
of particles per unit length}, along the stream.
Even though the previous sections show that the accelerations and orbits have converged to within a few
percent, we now investigate how well the stream density converges. 

\referee{\fig{stream_density_convergence} shows
the linear density along four streams using 100,000 particles each and traced along a sky projection as
seen by an observer at the galactic center
\waynerev{(so the unit along the length of stream is angular) at 6.7 Gyr}.
The four streams \waynerev{started with} identical progenitors in \referee{Orbit 1},
but in the smooth halos that are constructed using orders 10, 15, 20, and 25. Although the
\waynerev{overall profiles of the streams' densities on large scales are similar among
the four orders}, the detailed fluctuations are quite different. To investigate the significance of
these differences, we simulated a few additional streams with different random realizations of the
progenitor, but with the same number of particles, physical parameters, and orbit, in the same
smooth halo potential using order 10. We find that the small-scale differences between these 
additional streams are indistinguishable from the differences between the four panels
in \fig{stream_density_convergence}. Therefore, the differences shown in the figure are simply
due to stochastic noise of the particles, and our decomposition has sufficiently converged
at order 10.}


\referee{The convergence of our decomposition can also be seen in \fig{mass_loss},
which shows the mass loss for each
stream in \fig{stream_density_convergence}. Every large spike, or major mass loss \waynerev{event}, in each stream
is associated with a pericentric passage shown in the top panel of \fig{orbits_radial}. However,
the minor mass-loss events do not occur at the same rate among the four orders. Our additional
streams with different progenitor realizations as explained above show, again, that 
the small-scale differences for each decomposition order shown in \fig{mass_loss} are
indistinguishable from stochastic noise.}


\referee{We adapt order 10 for the smooth halo for the rest of our results. As we show in the next sections,
gaps caused by subhalos are longer than the spurious gaps caused by changing the decomposition order
as shown in \fig{stream_density_convergence}.} This is similar to the result of \citetalias{ngan14}, where the gap size 
distribution can be used to distinguish ``intrinsic'' gaps from subhalo gaps. In the future, we aim to repeat similar analyses
to \citetalias{ngan14} using various SCF decomposition orders for the realistic halo.}


\begin{figure}
    \includegraphics[width=3.5in]{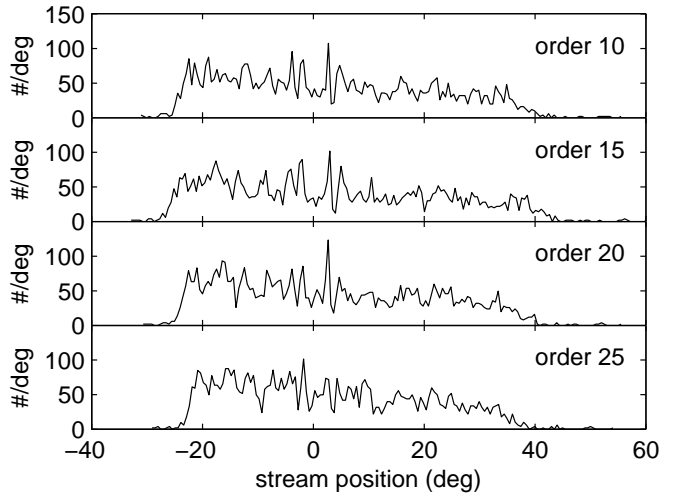}
    \caption{Four streams that start with the same initial condition\referee{s} with \referee{Orbit 1} in the smooth halo
    that is constructed using different decomposition orders. \waynerev{The streams' overall profiles on large scales
    are similar, but the \referee{density fluctuations on small scales} are different.}}
    \label{fig:stream_density_convergence}
\end{figure}

\begin{figure}
    \includegraphics[width=3.5in]{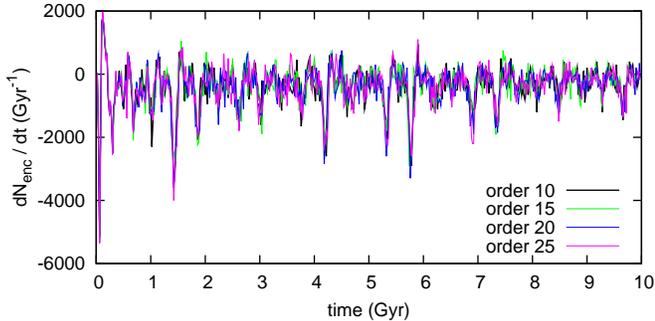}
    \caption{Rate of change in the number of particles enclosed in 0.1 kpc from the center of the progenitor
    as a function of time in \referee{Orbit 1}. Every spike
    in mass loss is associated with a pericentric passage.
    Even though the major mass loss events occur at the same time at the same rate among all orders,
    the minor events do not.}
    \label{fig:mass_loss}
\end{figure}

\subsection{\referee{Densities along the Streams}}

\subsubsection{\referee{Stream 1 in the Smooth Halo}}
\label{sec:planar_stream_smooth_halo}

\begin{figure}
	\includegraphics[width=3.5in]{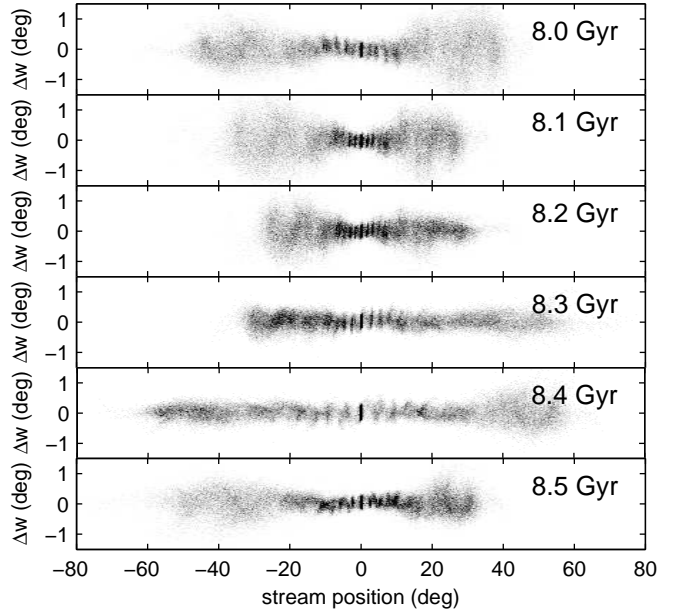}
	\caption{Sky-projected surface density of the stream in \referee{Orbit 1} in the smooth halo. The projection
	is seen from a hypothetical observer situated in the galactic center. Since the stream is extremely narrow
	compared to the size of its orbit, the curvature of the stream on the sky has been subtracted, and the
	progenitor has been shifted to have position at $0\degrees$. Note that the width of the stream
	($\Delta w$ on the y-axis) is about 50 times smaller than the length of the stream.}
	\label{fig:cartesian_inc0_sub0}
\end{figure}

\begin{figure}
	\includegraphics[width=3.5in]{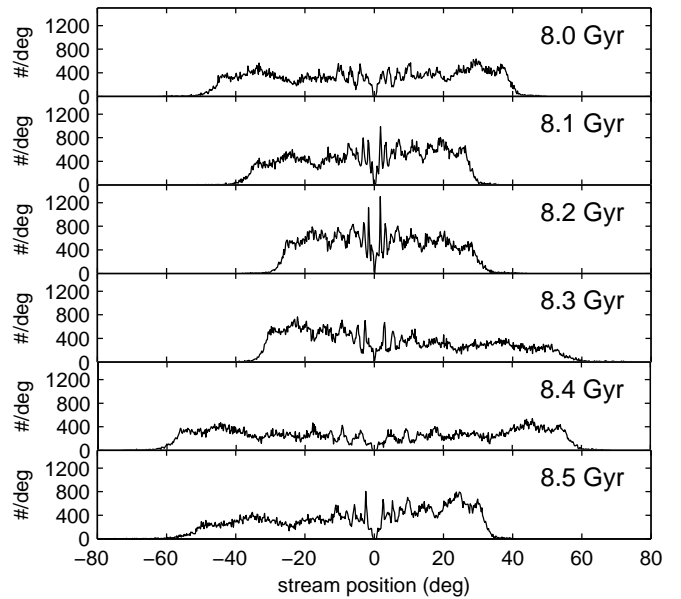}
	\caption{Linear density corresponding to the same stream shown in \fig{cartesian_inc0_sub0}. The
	progenitor at position $0\degrees$ has been masked out. \waynerev{Epicyclic overdensities appear as the
	first few spikes within about $10\degrees$ on either side of the progenitor (especially noticeable
	at 8.2 Gyr when the stream is compressed).}}
	\label{fig:density_inc0_sub0}
\end{figure}

We first examine the intrinsic features in \referee{the stream in Orbit 1 (hereafter ``Stream 1'')} in the smooth halo without any subhalos. 
\fig{cartesian_inc0_sub0} shows the surface density of the stream projected \waynerevv{on the sky from 8 to 8.5 Gyr}
as seen by an observer situated at the galactic center,
and \fig{density_inc0_sub0} shows the linear density along the stream at the same times. \waynerev{This
time frame roughly corresponds to one radial oscillation (though it is not a ``radial period''; see \fig{orbits_radial}),
so it allows us to investigate the 
streams' features as the streams stretch and compress during that oscillation.}

The resulting stream is long and narrow, with a length-to-width ratio of about 50 (which is slightly lower than
the observed values in GD-1 \citep{gd1}), so it can simply be represented by its linear density. The most prominent
feature in the linear density is the \waynerevvv{spikes located within $10^\circ$ away from either side of the progenitor. 
Upon inspection, we find that new spikes develop immediately after pericentric passages, and each spike originates
from the progenitor and migrates toward the ends of the stream. One} possible cause \waynerevv{of the density spikes} is EOs, \waynerev{which are caused by the ``piling up'' of
particles in their epicyclic orbits as they escape from the progenitor.} In the original derivations
in \citet{kupper08,kupper10}, the spacings between the EOs are calculated assuming an axisymmetric potential.
Although the VL-2 halo has no spatial symmetry, we can approximate it by \waynerev{spherically} averaging our potential model
(ie. taking $l_{max}=0$ in \eqn{phi}). Using a representative value of $R$ for the angular frequency $\Omega$
and epicyclic frequency $\kappa$, we can estimate the spacing
\begin{equation}
    |y_C|=\frac{4\pi\Omega}{\kappa} \left( 1 - \frac{4\Omega^2}{\kappa^2} \right) x_L, 
\end{equation}
where $x_L=GM/(4\Omega^2-\kappa^2)$, between the first EOs on either branch of the stream \waynerev{\citep{kupper08}}. For $R=24$ kpc
(roughly the galactocentric radius \waynerev{of the progenitor} at 8.3 Gyr), we obtain $\Omega\simeq8\,\mathrm{Gyr}^{-1}$ and
$\kappa\simeq12\,\mathrm{Gyr}^{-1}$. This gives $|y_C|\simeq0.77$ kpc, or $1.8^\circ$ when projected onto
the sky, which is almost a factor of 2 smaller than the measured spacing at $\sim3^\circ$ between the first two
density spikes at 8.3 Gyr. \waynerevvv{Note that as the stream compresses and stretches as it oscillates radially, the spacing
between the spikes can easily differ by factors of a few (eg. compare 8.2 and 8.4 Gyr in \fig{density_inc0_sub0}).}

The density profiles along the stream shown in \fig{density_inc0_sub0} indicate that \waynerevv{the density spikes near 
the progenitor} are very prominent inside the VL-2 potential and
are similar to those in \citetalias{ngan14} in a spherical potential -- they are regularly spaced and appear
very close to the progenitor. As we shall see in the next sections, gaps caused by subhalos are randomly
spaced and can occur anywhere along the stream.

\subsubsection{\referee{Stream 1 in the \waynerev{Lumpy} Halo}}

\begin{figure}
	\includegraphics[width=3.5in]{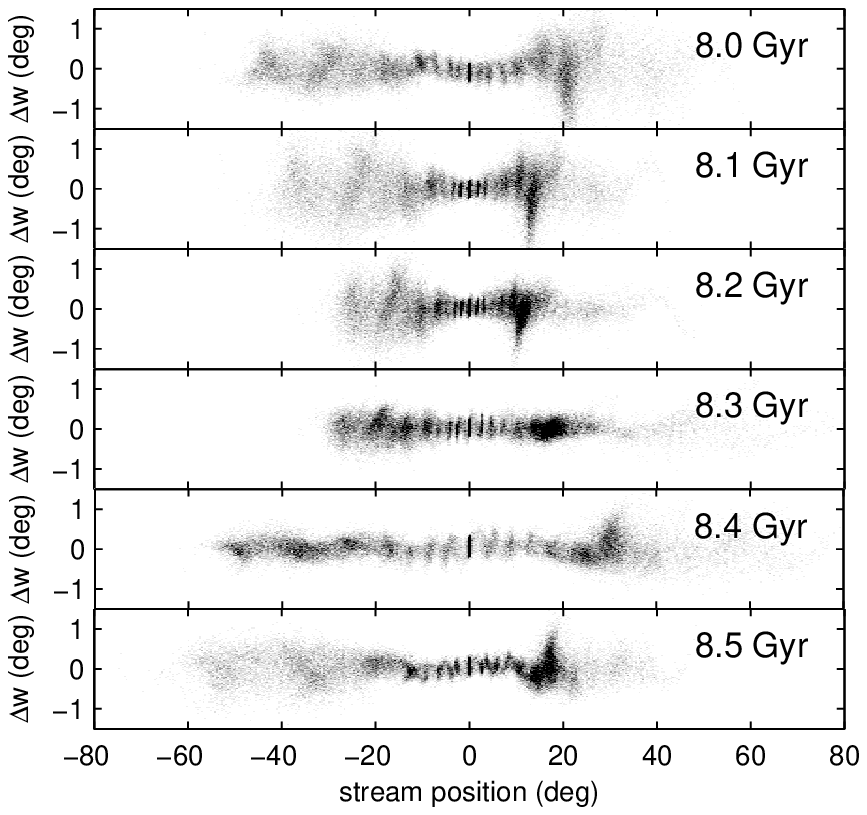}
	\caption{Same as \fig{cartesian_inc0_sub0}, but in a \waynerev{lumpy} halo with 3807 subhalos.}
	\label{fig:cartesian_inc0_sub3807}
\end{figure}

\figs{cartesian_inc0_sub3807}{density_inc0_sub3807} show the sky projection and linear density of \referee{Stream 1}
in a \waynerev{lumpy} halo with 3807 subhalos. Compared to the same stream in the smooth halo, the density clearly shows
gaps as local minima that are more pronounced and occur everywhere along the stream. The density spikes near the progenitor are still present
with similar spacing to the stream in the smooth halo. This is not surprising, since the extra subhalos contribute
only \waynerevvv{$\sim 1\%$} of the total mass of the smooth halo, \waynerevvv{so any effects due to the stream's orbit 
would not be different from the smooth halo case.}

Recall from Section \ref{sec:adding_subhalos} that our simulations in the \waynerev{lumpy} halo contains 3808 subhalos.
An interesting but rare event that occurred in the simulation with 3808 subhalos was that the progenitor had
a close encounter at less than one half-mass radius from a $1.8\sci{7}\Msun$ subhalo at an early time, which caused a sudden
burst in mass loss for the progenitor. This resulted in a stream that was dominated by two large but 
smooth spikes, one in each branch of the stream. When this subhalo was eliminated, we were able to resolve
the density fluctuations caused by
close encounters between the tidal tails and subhalos as shown in \figs{cartesian_inc0_sub3807}{density_inc0_sub3807},
rather than \waynerev{between} the progenitor and subhalos. In the future 
we aim to study the probability that a progenitor or its tidal stream becomes catastrophically perturbed
by subhalos in the VL-2 halo.

\begin{figure}
	\includegraphics[width=3.5in]{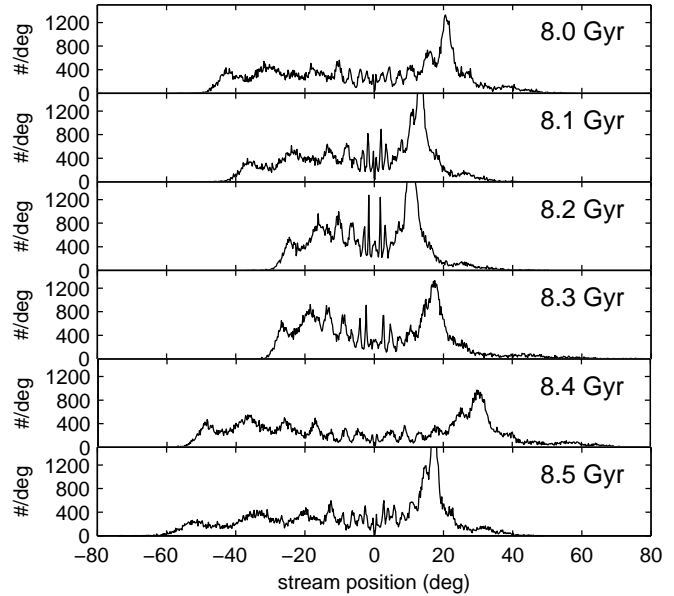}
	\caption{Linear density corresponding to the same stream shown in \fig{cartesian_inc0_sub3807}. Compared
	to the smooth halo case (\fig{density_inc0_sub0}), in a \waynerev{lumpy} halo the stream appears clumpier with more
	local minima in density everywhere in the tails of the stream.}
	\label{fig:density_inc0_sub3807}
\end{figure}

\subsubsection{\referee{Stream 2} in the Smooth Halo}

\begin{figure}
	\includegraphics[width=3.5in]{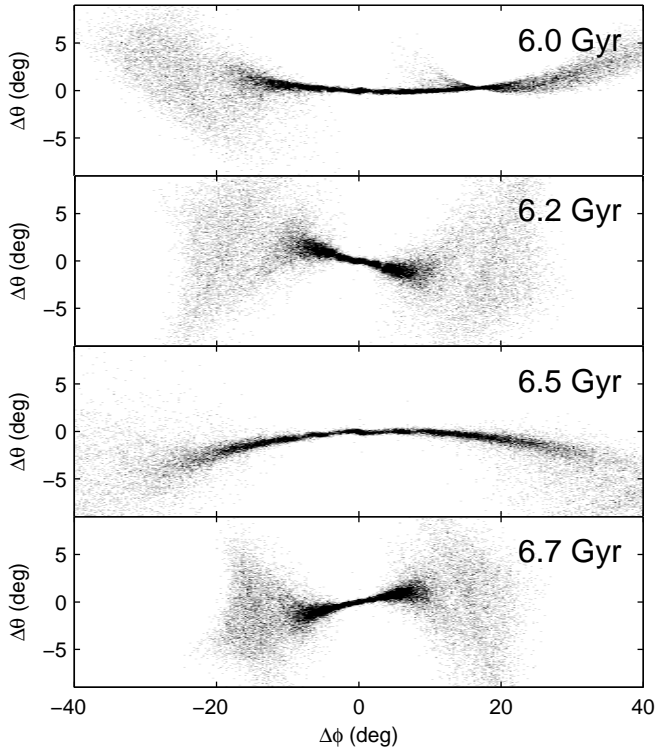}
	\caption{Sky-projected surface density of \referee{Stream 2} in the smooth halo. The projection
	is seen from a hypothetical observer situated in the galactic center. Compared to \fig{cartesian_inc0_sub0},
	the streams here contain diffuse ends sometimes as wide as the length of the stream. Rather than tracing
	a line and subtracting their curvatures on the sky, the streams shown here are simply shifted so that the
	progenitors have $\phi=\theta=0$, and then rotated so they appear roughly horizontal.}
	\label{fig:cartesian_inc70_sub0}
\end{figure}

\fig{cartesian_inc70_sub0} shows the sky-projected density of \referee{the stream in Orbit 2 (hereafter ``Stream 2'')} from 6.0 to 6.7 Gyr 
in the smooth halo. \waynerev{Similar to our investigation for \referee{Stream 1}, this time frame also covers roughly one radial oscillation
even though a well-defined radial period does not exist in the VL-2 potential (\fig{orbits_radial})}. Compared
to \referee{Stream 1}, \referee{Stream 2} becomes ``fluffy'' at the far ends of its tidal tails,
with widths sometimes comparable to the length of the narrow part of the stream. This is very different from  \referee{Stream 1}
shown in \fig{cartesian_inc0_sub0}, which has a high length-to-width ratio throughout.

\waynerevvv{An important implication is that
only the narrow segment of the stream may be dense enough to be observable, and the rest of the stream may be too diffuse.
Simulations in \citet{pearson14} in triaxial potentials also showed similarly diffuse features dubbed ``stream-fanning''
in the tips of tidal tails, which are not found in observations. The physical origin of stream-fanning has yet to be
understood and is beyond the scope of our study, so it is not clear whether our \referee{Stream 2} is exhibiting
stream-fanning.}

\begin{figure}
	\includegraphics[width=3.5in]{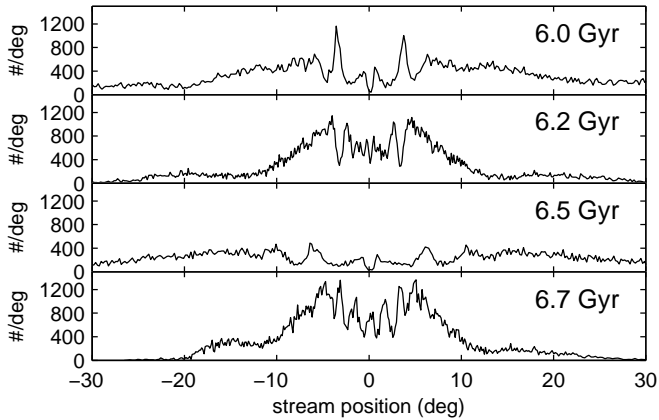}
	\caption{Linear density corresponding to the same stream shown in \fig{cartesian_inc70_sub0}. A
	line has only been traced along the regions up to $20\degrees$ away from the progenitor.  The progenitor
	at position $0\degrees$ has been masked out. The density profiles show prominent \waynerevvv{density spikes
	near the prongenitor, similar to \referee{Stream 1}}, but the profiles quickly fall
	off to the diffuse regions without many intrinsic features.}
	\label{fig:density_inc70_sub0}
\end{figure}

Because of the diffuse tails, it is no longer appropriate to analyze the entire stream simply by the linear density.
Nevertheless, we can trace the linear density along the narrow part of the stream out to about $10^\circ$--$20^\circ$ 
away from the progenitor. \fig{density_inc70_sub0} shows the densities along the stream at each time shown in
\fig{cartesian_inc70_sub0}. The density profiles show prominent \waynerevvv{density spikes} next to the progenitor, and
the spacings \waynerevvv{between those spikes}
are in rough agreement with the derivation in Section \ref{sec:planar_stream_smooth_halo}. Other than this,
the density profiles fall off very quickly and smoothly toward the diffuse ends of the stream.

\subsubsection{\referee{Stream 2} in the \waynerev{Lumpy} Halo}

\begin{figure}
	\includegraphics[width=3.5in]{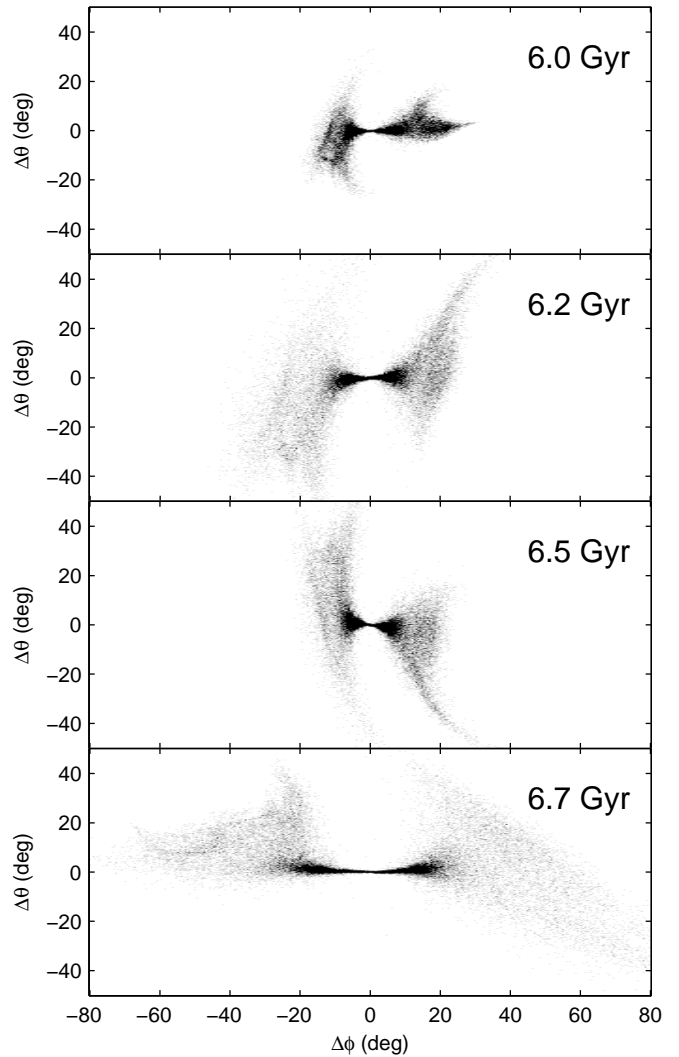}
	\caption{\waynerevvv{Same as \fig{cartesian_inc70_sub0},} but in the \waynerev{lumpy} halo. The stream
	overall appears similar to the case in the smooth halo, but the diffuse ends of the stream are even
	more prominent, and sometimes even wider than the narrow part of the stream itself.}
    \label{fig:cartesian_inc70_sub3808}
\end{figure}

\fig{cartesian_inc70_sub3808} shows the sky-projected density of \referee{Stream 2} from 6.0 to 6.7 Gyr
in a \waynerev{lumpy} halo with 3808 subhalos. The stream resembles the case in the smooth halo
(\fig{cartesian_inc70_sub0}) with very fluffy ends, but in a \waynerev{lumpy} halo they are even more prominent.
In the fluffy regions the width can exceed the length of the narrow regions, which span only 
$10^\circ$--$20^\circ$.

\begin{figure}
	\includegraphics[width=3.5in]{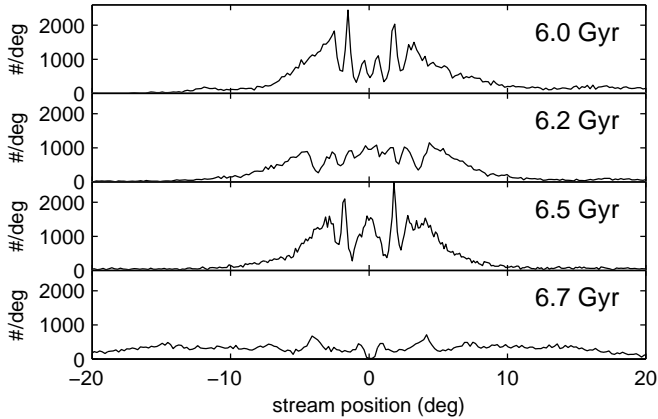}
	\caption{Linear density corresponding to the same stream shown in \fig{cartesian_inc70_sub3808}.
	Similar to the case in the smooth halo, linear density is only appropriate at the regions near
	the progenitor where the stream is narrow. Aside from \waynerevvv{the density spikes} near the progenitor, the density
	profile is smooth and featureless.}
	\label{fig:density_inc70_sub3808}
\end{figure}

The linear density along the stream can be plotted, but similar to the case in the smooth halo, it
is only appropriate to the narrow regions near the progenitor. As seen in \fig{density_inc70_sub3808},
the density profiles in those regions are also smooth with the exception of the first or second \waynerevvv{density spikes}.
Therefore, for this stream, there are no distinguishable signatures in the linear density that traces
the existence of subhalos.

Perhaps a more interesting aspect of this stream is the tips of its tidal tails. The wide and diffuse 
tails shown in \fig{cartesian_inc70_sub3808} are reminiscent of ``shell''-like features found in 
satellite systems that plunge near the galactic center on very eccentric orbits. With apo- and pericentric
distances at $r_a\simeq30$ kpc and $r_p\simeq17$ kpc, respectively, the stream's eccentricity is
$e=(r_a-r_p)/(r_a+r_p)\simeq0.3$. This implies that satellites do not need to be on eccentric orbits
to have significant parts of their tidal tails disrupted such that their surface brightnesses may be
even lower than that of the narrow stream itself, which is already difficult to observe. 
\waynerevvv{In \citetalias{ngan14}, the same progenitor orbiting a spherical and idealized halo with similar
eccentricity to \referee{Streams 1 and 2} here could only produce a narrow \referee{stream}. This indicates that using 
a realistic halo is important when studying the effects that subhalos have on GD-1-like streams.} In a future
study, we aim to simulate the tidal disruptions of \waynerevvv{globular-cluster type} satellites in many orbits
in both the \waynerevvv{smooth and} \waynerev{lumpy} halos in order to study the survival rate of tidal tails.

\section{Summary and conclusion}
\label{sec:summary_and_conclusion}

We constructed \referee{potential models} using the high-resolution dark matter halo in the $z=0$
snapshot of the Via Lactea II
(VL-2) simulation. In the high-resolution ``main'' halo, we used a halo finder to remove and isolate subhalo particles.
\waynerev{The potentials of the main halo and the subhalos were then} constructed individually. This allowed us to simulate
tidal streams in the main halo with and without subhalos. We investigated the difference in the stream between
the two cases and showed that, even in a realistic potential without using idealized profiles, streams remain a valuable probe
\waynerev{to detect subhalos whose existence is a crucial prediction of the \lcdm\ model}.

The SCF method \citep{scf} has previously been applied to the Aquarius simulations
\citep{aquarius, hex} as a tool to study dynamics inside existing high-resolution simulations of
dark matter halos. For the first time, we applied it to the VL-2 simulation to examine the 
subhalos' dynamical influence on tidal streams. The SCF method is a powerful and parallelizable 
method that efficiently computes the potential to arbitrary accuracy using a set of complete and 
bi-orthonormal basis functions.

The potentials in the main halo and all the subhalos were constructed using the SCF method. When the subhalo
particles were removed from the main halo, the main halo's potential was constructed using an order 10 decomposition
to ensure that the main halo remained smooth \waynerev{and contained no granularities due to the removal of its
subhalos}.
After the subhalos' individual potentials were also constructed, they were added to the \waynerev{smooth} halo in orbits with
the initial positions and velocities as found by the halo finder. This
eliminated the need for any assumptions for the profiles and distributions for the subhalos in the \waynerev{lumpy} halo.

We simulated the tidal disruption of a star cluster as an N-body system that used the main halo's and
subhalos' potentials as external forces. The cluster followed two orbits, one with initial velocity
\referee{with zero $z$ component (``Stream 1'')}, and another with the same speed but inclined from the $xy$-plane by $70^\circ$
(\referee{``Stream 2''}). In a 
potential without spatial symmetry, neither orbit was confined to any orbital planes, but both orbits
were chosen to have apo- and pericentric distances at about 30 and 17 kpc, respectively. These values
are comparable to the \waynerev{inferred} orbit of GD-1 from observations \citep{gd1,willett09}. We simulated
each stream in both the smooth halo and the lumpy halo.

\referee{Stream 1} remained narrow for 10 Gyr, with length-to-width ratio $\simeq 50$,
which was comparable to the observations of GD-1. Even though the smooth halo had
no spatial symmetry, we still found \waynerevvv{density spikes that were similar to} EOs
near the stream progenitor. 
In the smooth halo, there were density fluctuations along the stream that were not found in a
spherical halo, but the fluctuations were not as prominent as the ones found in the stream in the
same orbit but in a \waynerev{lumpy} halo. \waynerevv{In the future we aim to repeat analyses similar to
\citet{ngan14}, which used gap size distributions to distinguish ``intrinsic'' gaps and subhalo gaps in 
streams. Since gap size distributions are observable \citep{gd1gaps}, understanding them for simulated streams
in a realistic halo is an essential step for comparing simulations and observations.}


\referee{Stream 2} was narrow only up to $10^\circ$--$20^\circ$ away from the progenitor
as seen by a hypothetical observer situated at the galactic center. Further
\waynerev{along the stream} from the narrow part, the stream developed fluffy 
features that were almost as wide as, if not wider than, than the length of the narrow part of the stream.
We found that density spikes, \waynerevvv{similar to the ones in \referee{Stream 1},} near the progenitor
dominated the linear density along the narrow part without any distinguishable
signatures of subhalos. However, in the \waynerev{lumpy} halo, the stream featured wider and even more diffuse 
tails than in the smooth halo. This means that in the \waynerev{lumpy} halo the stream may be even more
difficult to observe than in the smooth halo.

Dynamics in the VL-2 halo potential is beyond the scope of this study. In this study we used only two
streams in two arbitrary but similar orbits to illustrate the VL-2 halo's complexity compared to idealized
and spherical halo models that were used in previous studies \citep{yoon11, carlberg13, ngan14}.
\waynerevvv{In a spherical halo, our two streams in this study would appear identical and would not reveal
the complicated morphologies especially demonstrated by \referee{Stream 2}.}
In the future, we aim to use this VL-2 model to simulate more streams in more orbits and to perform much
more detailed analysis to each stream.

\waynerev{B.B., R.W., and A.S. are supported by NSF grant OIA-1124403, and P.M. by OIA-1124452.
We thank the anonymous referee for a constructive report. W.N. also thanks John Dubinski for
useful discussions for the SCF method.
Computations were performed on the GPC supercomputer at the SciNet HPC Consortium. SciNet is funded by
the Canada Foundation for Innovation under the auspices of Compute Canada; the Government of Ontario;
Ontario Research Fund - Research Excellence; and the University of Toronto.}

\bibliographystyle{apj}

\bibliography{VLII}




\end{document}